# Future Decline of African Dust: Insights from the Recent Past and Paleo-records

Tianle Yuan[1,2], Hongbin Yu[1], Mian Chin[1], Lorraine A. Remer[2]

[1]Earth Sciences Division, NASA Goddard Space Flight Center, Greenbelt, Maryland

[2]Joint Center for Earth Systems Technology, University of Maryland, at Baltimore County, Baltimore, Maryland

**Abstract**

African dust is a major aerosol source and interacts with diverse aspects of the climate system such as atmospheric composition and energy balance[1,2], tropical Atlantic sea surface temperature (SST)[3], nutrient cycle to terrestrial and marine ecosystems[4], and cloud and precipitation[5]. African dust exhibits strong variability on a range of time scales. Here we show that the interhemispheric contrast in Atlantic SST (ICAS) drives African dust variability on interannual, multidecadal, and millennial timescales, and a strong anthropogenic decline of African dust in the future can be expected due to the projected increase of the ICAS. During the recent past, the ICAS is found to significantly correlate with various dust observations and proxies that extend as far back as 1851. Physically, positive ICAS anomalies induce large-scale circulation changes that push the Intertropical Convergence Zone (ITCZ) northward and decrease surface wind speed over African dust source regions, which reduces dust emission and transport to the tropical Atlantic. The ICAS-ITCZ-dust relationship also finds robust support from paleo-climate observations that span the last 17,000 years. The ICAS drive of African dust variability is consistent with documented relationships between dust activity and Sahel precipitation[6], the North Atlantic Oscillation[7], and time series of a surface wind speed pattern over Northern Africa[8], and offers a unified framework to understand them. The ICAS-dust connection implies that human activities that change ICAS through emitting greenhouse gases and pollutions have affected and will continue to affect African dust. Climate models project that anthropogenic increase of the ICAS can push the ICAS value to surpass the highest level attained during the Holocene by the end of this century and decrease African dust activity by as much as 60% off its current level, which has broad consequences for aspects of the climate in the North Atlantic region and beyond.

**Text**

The inter-hemispheric temperature contrast, averaged over a longitude span wide enough such as an ocean basin, emerges as an important variable for the atmospheric energy balance and related dynamics [9-11]. Perturbation of the temperature contrast leads to an adjustment of the strength of Hadley cells in two hemispheres to restore energy balance. The Hadley cell in the relatively warming hemisphere slows down while its counterpart in the other hemisphere speeds up in order to transport more energy towards the relatively cooling hemisphere and reduce the temperature contrast [9-11] (Fig. 1A for an illustration of northern hemisphere warming). The lower branches of both Hadley cells, responsible for the trade winds, converge and give rise to the intertropical convergence zone (ITCZ). Slowing down of the Hadley cell in the relatively warming hemisphere reduces surface wind speed in the same hemisphere while increases surface wind speed in the other hemisphere. The lower branch of the Hadley cell in the Northern Hemisphere covers most African dust source area where the surface wind speed is a key driver for African dust activity [1,2,12]. At the same time, the latitudinal position of the ITCZ is pushed towards the warming hemisphere, affecting the transport and removal of dust.

The energy balance view therefore offers testable predictions on the interplay among the Inter-hemisphere Contrast in Atlantic Sea surface temperature (SST), or ICAS, African dust activity and the ITCZ position over the Atlantic Ocean over interannual and longer time scales. First, we define the ICAS as the area-weighted SST difference between the North (-80°E~20°E, 0°N~70°N) and South Atlantic (-70°E-20°E, 70°S~0°S) Oceans. Positive ICAS perturbations shall be associated with northward movement of the ITCZ, reduced surface wind speed in the African dust source regions, and thus reduction in African dust [Fig. 1A]. We illustrate the interplay in a large-scale context with analysis of spatially resolved surface wind from reanalysis data[13] and precipitation data from the Global Precipitation Climatology Project (Fig. 1B). Fig. 1B shows regression maps between the ICAS and the anomalies of precipitation and the surface wind vectors for the boreal summer season of June, July, August (JJA). Over the Saharan dust source regions, there is a clear shift of wind direction to be more southwesterly, which is an over-land extension of a large-scale, systematic southwesterly surface wind anomaly covering the tropical North Atlantic. The southwesterly surface wind anomaly is opposite in direction to that of the climatological northeasterly over the dust source regions. It therefore decreases the surface wind speed over much of the dust source regions and suppresses dust emission as well as transport to the tropical North Atlantic[1,14]. For precipitation, a longitudinal band of positive precipitation anomaly over the Sahel region is associated with positive ICAS. It is accompanied by a narrower and more longitudinally confined band of negative precipitation anomaly to the south, along the northern coast of the Gulf of Guinea. These two bands of precipitation anomaly with alternating signs suggest a northward movement of the intertropical convergence zone (ITCZ) associated with a positive ICAS. These observations are therefore consistent with our hypothesis based on the energy balance view [Fig. 1A].

Empirical evidence of the recent 160 years supports a close relationship between ICAS and African dust variability as well as the ITCZ position. Figure 2 shows various time series of African dust activity together with the ICAS as well as Sahel precipitation[15] and the Atlantic Multidecadal Oscillation (AMO) index[16]. Dust activity measurements and proxies include a synergy of satellite and coral dust proxy data[17], which is referred to as "the Cape Verde data" for short, proxies based on time variations of a surface wind speed pattern[8] from the European

Center for Medium-Range Weather Forecasts Interim reanalysis (ERA-I) [13] and NOAA-CIRES 20th Century Reanalysis[18] (CIRES-20CR), "the ERA-I data" and "the CIRES-20CR data" for short, respectively, and the long-term Barbados surface dust concentration measurements[6], "the Barbados data". The ERA-I data [Fig. 2a] indicates that dust activity experienced an overall decreasing trend since the beginning of the 1980's and it is well-correlated with the ICAS time series (r = -0.64, p< 0.001). The Cape Verde data [Fig. 2c] extends Atlantic dust activity measurements to 1955. It shows an increasing trend between 1950s and the end of 1970s and a decreasing trend afterwards, with the latter half agreeing with the ERA-I data. The Barbados record [Fig. 2b] has a similar trend since the 1950s, but its minimum in the 1960s dips lower than the Cape Verde data. Correlations between dust data sets and the ICAS are statistically significant at 99.9% (r =-0.63, p<0.001) and 95% level (r= -0.3, p<0.05) for the Cape Verde and the Barbados data, respectively. For the Barbados record, the correlation coefficient increases to -0.58 if the raw annual data is low-pass filtered (0.1 $yr^{-1}$ is the cut-off), which may suggest that point measurements at Barbados have a strong interannual variability component that is not directly related to the ICAS. The CIRES-20CR data [Fig.2e] spans 162 years from 1851 to 2012. During this period, the African dust activity experiences roughly two oscillation cycles. After a brief period of slightly decreasing trend, dust activity increases from 1870s to 1910s, persists until a quick drop at the beginning of 1950s, then follows similar trajectory as the Cape Verde data and the Barbados data. The ICAS also undergoes two cycles during this period, but with opposite phase [Fig. 2f]. The correlation coefficient between the ICAS and the CIRES-20CR data is -0.21 (P <0.01). Prior to about 1900's SST measurements over the South Atlantic Ocean is quite sparse[19], which may affect the reliability of calculated ICAS during this period. The AMO index, which is based on North Atlantic SST measurements, can be used as a proxy of the ICAS [20,21] before 1900's if we assume the high correlation between them after 1900's (r = 0.72, p<0.001) holds. Using the AMO index as proxy, the correlation coefficient between ICAS and the CIRES-20CR data increases to -0.4 (p<0.001). Low-pass filtered time series (thick black lines in each panel) show that dust data and the ICAS undergo phase changes (indicated by vertical lines in Figure 2) around similar time, further clarifying the ICAS-dust connection. Positive ICAS is also strongly correlated with the Sahel precipitation index (SPI) (r=0.48, p<0.001) as shown in Fig. 2d. This is consistent with a northward movement of ITCZ position associated with the positive ICAS (Fig. 1) because northward displacement of ITCZ brings more precipitation into the Sahel region (Fig. 1).

The ICAS-ITCZ-dust relationships in Figs 1 and 2 find further support from paleo-records of the ICAS, African dust activity, and the ITCZ position in the past nearly 17,000 years. The ICAS during this period is approximated by the hemispheric temperature difference based on temperature reconstructions [22,23] (see Data and Method). The Atlantic ITCZ latitudinal position is shown to positively correlate with the Titanium concentration record from the Cariaco Basin[24] (also see Data and Method). African dust activity is approximated by terrigenous flux and dust fluxes from three locations in the Atlantic Ocean: dust fluxes from two cores around the Bahamas (26°N, 78°W), terrigenous flux at the Ocean Drilling Program site 658C (ODP658C, 20.75 °N; 18.58 °W) off the Mauritania coast, and dust flux measured at a tropical Atlantic Ocean site (5.33°N, 33.03°W). The ICAS undergoes a general upward trend between approximately 20,000BP to 6,000BP and a gradual downward trend since 6,000 BP, punctuated by several distinct events (Figure 3). These events together with the gradual trends are used to test the ICAS-dust-ITCZ connections.

Between 14,000 and 17000 years BP, the northern hemisphere differentially cools as the thermohaline circulation slows down[25,26] and the ICAS is at its minimum [Fig. 3]. Concurrently, dust records from three locations all show an elevation of African dust activity, although the ODP658C record only covers small part of this period. Between 12,700 and 14,700 BP, the northern hemisphere warms relatively to the southern hemisphere[27], increasing the ICAS. Dust activity decreases, compared to the previous period, with the increasing ICAS in all three dust records though during the latter part of this period, the dust increases at the ODP658C site. The position of the ITCZ moves northward as indicated by the increasing Titanium concentration. Both dust changes and the ITCZ movement are consistent with theoretical predictions. During the Younger Dryas period, between 11,700 and 12,700 BP, the northern hemisphere rapidly cools relatively to the southern hemisphere and the ICAS decreases sharply. While the ITCZ moves significantly southward as expected (Fig. 3a), the strong decrease of ICAS is only accompanied by strong initial dust increase at the African coast site ODP658C. At Bahamas sites that are further north (~26°N) and away from the West African Coast, dust activity decreases significantly. The differing responses at the ODP658C and the Bahamas sites may be explained if the southward shift of the ITCZ (Figure 3a) pushes the main dust transport route southward by an extent that is large enough to cut off the dust transport to the Bahamas sites. Starting around 11,700 years BP, the northern hemisphere warms up relatively to the southern hemisphere quickly and the ITCZ moves northward. During this period, dust activities at all three sites decrease. Dust activities from all sites reach a global minimum during the African Humid Period[28], roughly 10,000 to 5,000 BP, when the ICAS peaks and the ITCZ reaches the northernmost position. The greening of the Sahara[28] during this period may have also contribute to the African dust minimum. Towards the end of the African Humid Period, the ITCZ shows a slight southward movement while the dust activity at the ODP658C site increases significantly, which is however not evident over other sites. The ICAS decreases gradually without drastic changes, which may suggest that the dust increase at the African margin site could be a more localized response for reasons that are not clearly understood yet. The ICAS then gradually decreases and the ITCZ moves southward gradually, while dust activities from all sites generally increase. During the relatively brief 'Little Ice Age', the decreased ICAS and southward ITCZ shift are accompanied by an increase of dust activity at the ODP658C site and one site at the Bahamas.

Regression analyses further support these interpretations (Figure 4). The ICAS proxy data for the recent 11,300 years are used for this analysis and similar trends are found for the period between 6,000BP and 17,000BP (see Fig. S2). Both the ICAS proxy and the ITCZ position proxy data are averaged to match the time resolution of the terrigenous flux data at the African margin site (ODP658C). The ICAS proxy is strongly anti-correlated with the dust activity ($r=-0.93$, $p<0.001$) and positively correlated with the Titanium concentration, thus the ITCZ position ($r=0.95$, $p<0.001$). Both the ITCZ position and dust activity proxy data seem to have two regimes for this period, with a separation at the 5,000BP mark in time, which is most obvious for dust activity jump in Fig. 3b. However, the slopes between the ICAS and the ITCZ and between the ICAS and dust activity are very similar for the two regimes, which indicates robust sensitivity of the ITCZ and dust activity to the ICAS.

Our hypothesis on the interplay among dust activity, the ICAS, and the ITCZ position finds not only strong support in observations from both recent and distant past, but also provides a robust large-scale framework to understand African dust variability on decadal or longer timescales. This framework provides a unified way to understand many previous studies that deal with the driver of African dust variability. For example, the reported correlation between Sahel precipitation and dust variability[6] can be understood as a direct result of the ICAS simultaneously driving changes in Sahel precipitation (Figure 1d and Figure 2) and dust activity. The Sahel sits on the northern limit of the ITCZ over the continent and receives its precipitation from the ITCZ excursions. Positive (negative) ICAS anomalies pull the ITCZ northward (southward) and increase (decrease) precipitation in the Sahel region (Figures 1, 3, and 4). Meanwhile, positive ICAS anomalies reduce dust emission by slowing down surface wind speed (Figure 1 and 2), thus creating an anti-correlation between the Sahel precipitation and dust activity. This interpretation is also in accordance with multiple studies that show surface wind speed in dust source region, instead of new dust sources in the Sahel due to dryness and/or reduced vegetation cover, as the dominant factor for African dust variability in recent decades and distant past [1,2,29]. The framework is also consistent with the relationship between the wintertime North Atlantic Oscillation (NAO) and dust in wintertime because the NAO is closely related to the AMO/ICAS[30]. The close relationship between the ICAS and the position of ITCZ also agree with the role of ITCZ in affecting African dust transport. It is worth noting that this framework is most applicable to decadal and longer time scales while other higher frequency processes such as the El Nino Southern Oscillation[31] can still play important roles at shorter time scales.

The robust relationship between the ICAS and dust variability as supported by the recent and distant past makes it a useful predictor to project future changes of African dust in the context of climate change. Using ICAS as a predictor has certain advantages over previously used variables such as the Sahel precipitation, the NAO, and surface wind speed pattern. This is because regional changes of these variables simulated by general circulation models (GCMs) are uncertain, even when forced with observed boundary conditions. For example, under similar boundary conditions and forcing, surface wind speed in the dust source region shows strikingly different changes in current GCMs[14]. Temperature on the other hand can be more reliably predicted based on forcing scenarios. Assuming that the same relationship between the ICAS and dust activity operates in the future, changes in dust may be more reliably predicted using model generated SST fields. We calculate the ICAS for a set of models participating in Phase 5 of the Coupled Model Intercomparison Project Phase 5 (CMIP5) under two forcing scenarios: Representative Concentration Pathway (RCP) 8.5, and RCP4.5. In both scenarios, the multi-model mean of the ICAS is projected to increase significantly, pointing to a substantial reduction in African dust activity in the future (Figure 5). In particular, the multi-model mean ICAS predicts that the African dust activity will be two standard deviations below the mean of the past 160 years in about three and five decades for RCP 8.5 and RCP4.5 scenarios, respectively, which would cut the current dust loading over the tropical Atlantic by about 30%. In the RCP8.5 scenario, dust activity will decrease to this level by 2050 and continue to decline as the ICAS increases due to continued anthropogenic warming. Dust activity is projected to decrease to that level sooner if only models that better simulate past AMO[32] are included (see Data and Methods for models). The projected reduction in African dust by the end of this century using this set of models is about twice of the all model mean (right panels of Fig. 5) and African dust loading over the Tropical Atlantic will only be at 40% of its current level, about 20% of the level in the 1980's. In fact, for this set of models the multi-model mean ICAS will approach and even surpass the African Humid Period peak by the end of this century, which would imply a new global minimum of African dust activity for the last 17,000 years. Decadal to multidecadal natural variability such as the AMO could alleviate or exacerbate the dust decline depending on its phase[14]. For the same reason, radiative forcing induced by human activities in the past may have affected ICAS and thus dust activity[33].

The strong decline of African dust due to future ICAS increase will improve air quality in downwind areas such as countries around the Mediterranean and Caribbean Basin. Transported dust over the tropical North Atlantic has a strong cooling effect on the ocean surface via direct and indirect radiative effects[3]. The strong decline in African dust over the tropical North Atlantic constitutes a positive dust feedback to the anthropogenic warming of the ICAS since dust differentially cools the tropical North Atlantic SST and has minimum impact on the South Atlantic, which may further reduce the African dust loading[14]. The tropical North Atlantic SST will experience a unique additional warming due to the strong decline in dust compared to other tropical oceans, which has implications for zonal overturning circulation. The tropical Atlantic warming can have wide range of effects such as accelerating Antarctic sea ice loss[34] and inducing more hurricane activity in the Atlantic basin[21]. Reduced African dust emission and transport will decrease the supply of nutrients such as iron and phosphorus to ecosystems in the Atlantic Ocean and Amazon forests[4], affecting the biogeochemical cycles. Lastly, our results underscore that continued monitoring of dust activity from space for the next 30 years will be critical for detailed understanding of the African dust variability in a warming world.

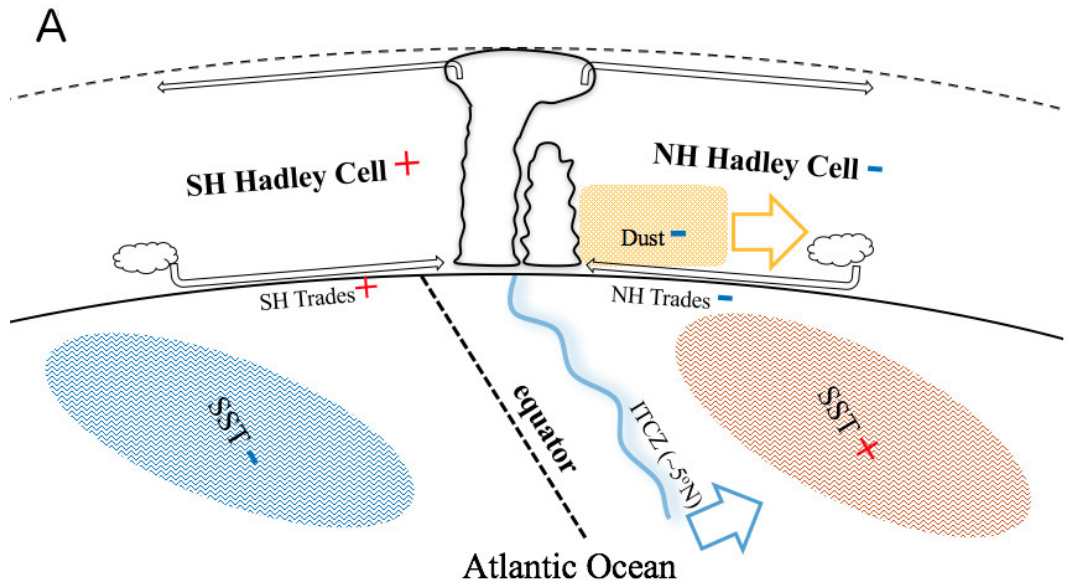
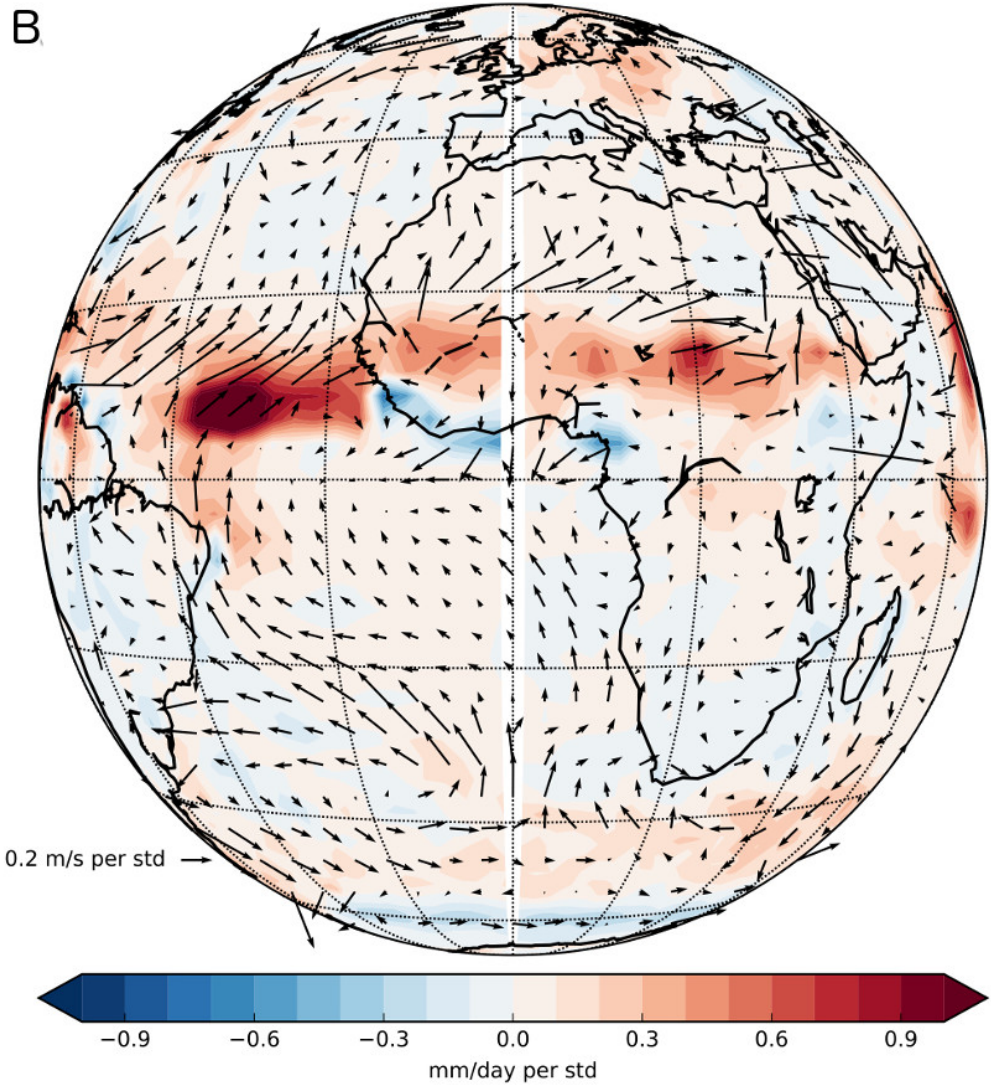

Figure 1: A) A schematic of the theoretical framework proposed to explain the relationships among the ICAS, the ITCZ position, and African dust. In current climate, the Atlantic ITCZ sits around 5°N in terms of annual mean with the main dust transport route to its north (about 12°N, see Fig. S1). With increasing ICAS, the northern hemisphere (NH) Hadley cell weakens and associated NH trade wind speed decreases, reducing dust emission and transport from Africa. Meanwhile, the ITCZ moves northward (light blue arrow) together with the dust transport route (light yellow arrow). B) Regression of ICAS against precipitation and surface wind during JJA between 1979 and 2015. Color shaded areas and wind vectors depict respectively precipitation and surface wind anomalies associated with one standard deviation of ICAS. It indicates a northward displacement of the ITCZ and strengthening and weakening of the trade winds to the south and north of the ITCZ, respectively because prevailing trade winds are northeasterly and southeasterly in the northern and southern hemispheres, respectively.

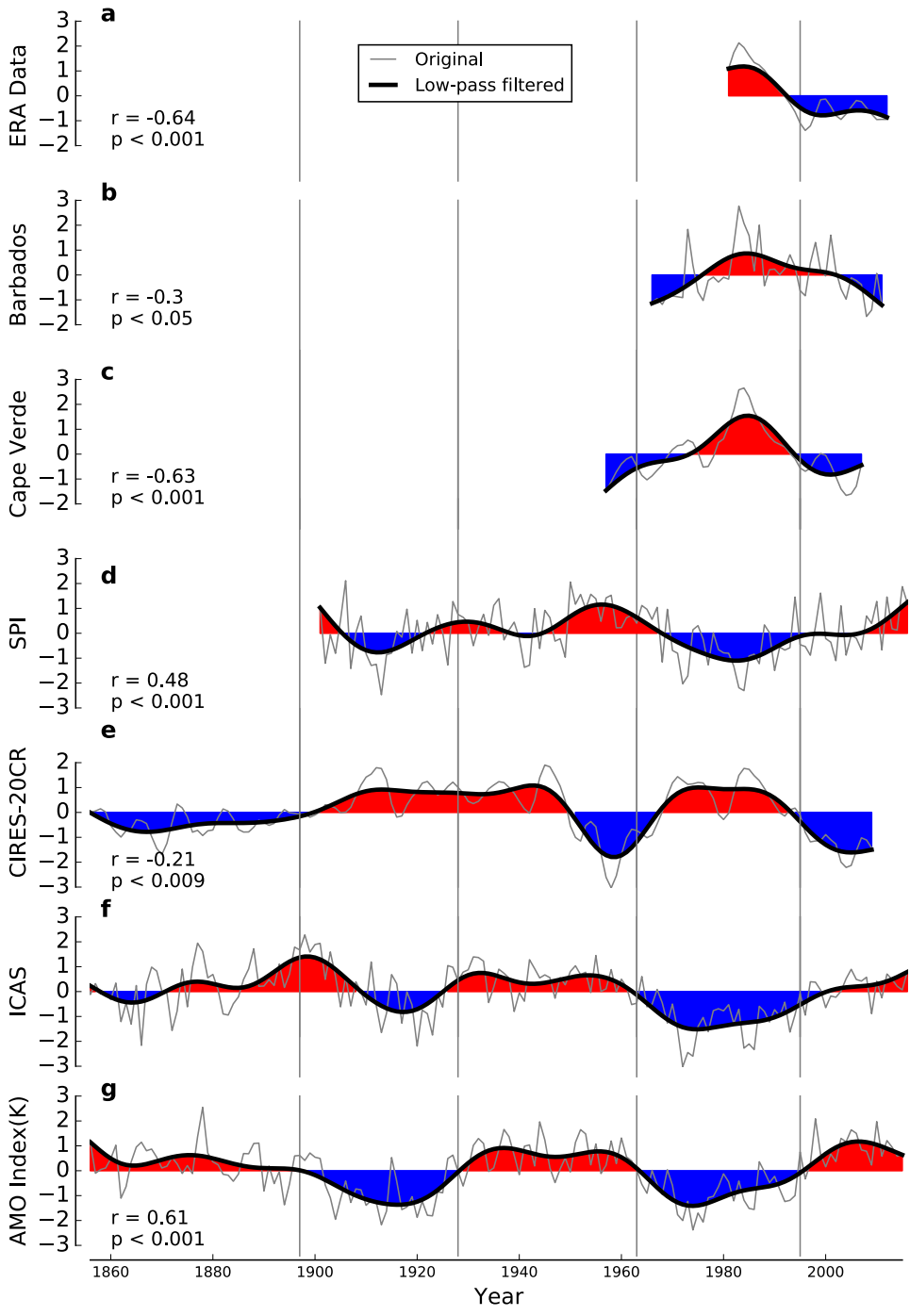

Figure 2: Time series of various dust activity proxies in the recent past, with vertical lines marking AMO phase changes. Both the original (grey lines) and low-pass (0.1 yr$^{-1}$ is the cut-off) filtered (bold solid) dust records are plotted. Each time series is detrended and then normalized by its standard deviation. Red (blue) shading indicates positive (negative) values. a) dust proxy based ERA-interim wind[8]; b) dust time series from Barbados record[6]; c) dust proxy based on Cape Verde record[17]; d) Sahel precipitation index[15]; e) dust proxy based on CIRES-20CR wind[8]; f) ICAS calculated based on the Hadley Centre Sea Ice and Sea Surface Temperature (HadISST)[19] data since 1850; g) AMO index since 1856[16]. The correlation coefficients and associated p values are shown for each variable when regressed against the ICAS data using raw annual data.

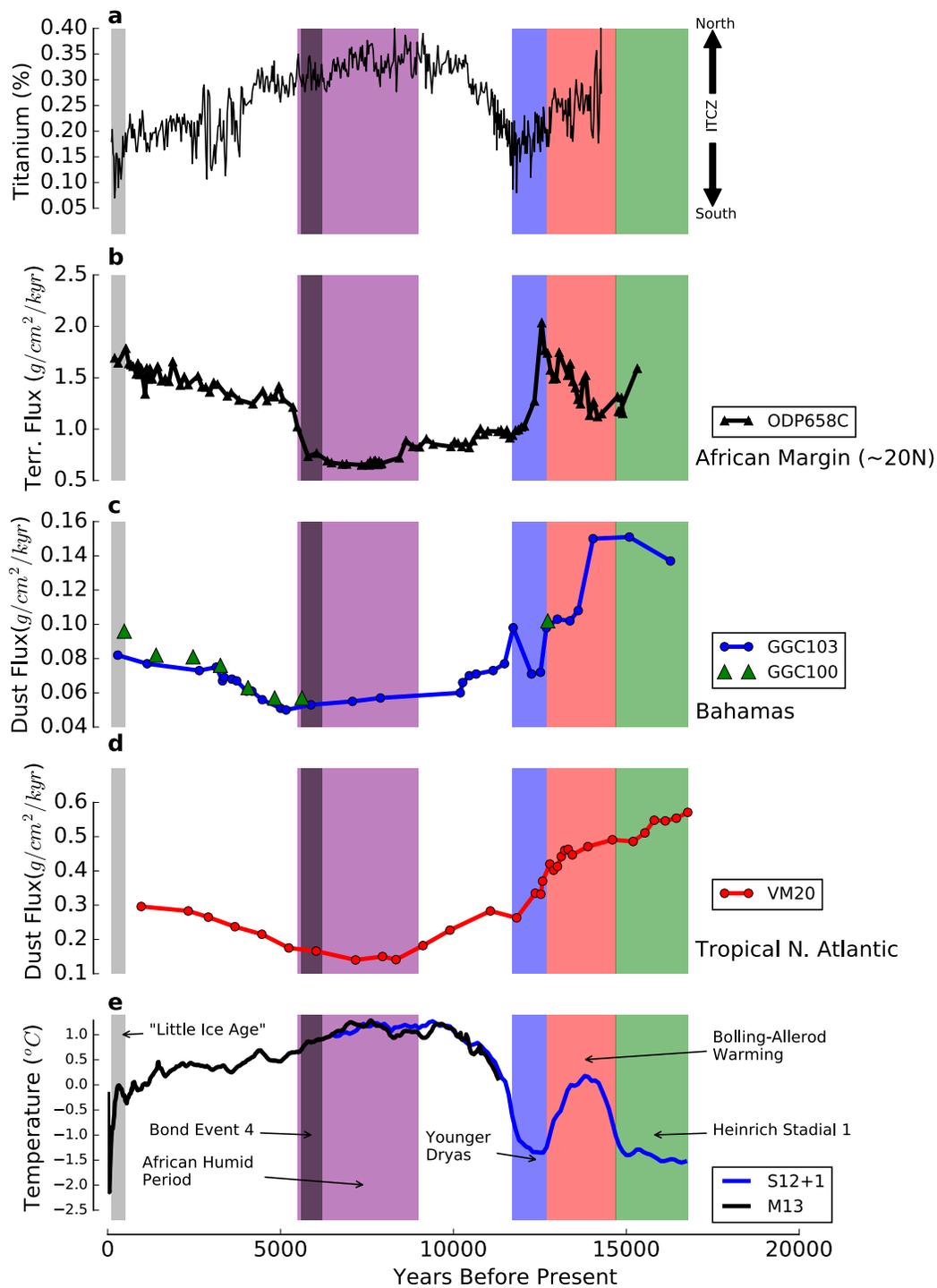

Figure 3: a) Titanium concentration from the Cariaco Basin, a proxy for ITCZ meridional position, for the last 14,000 years. Higher values indicate more northward position of the ITCZ. b) Terrigenous flux at the Ocean Drilling Program site 658C (20.75 °N; 18.58 °W), a proxy for dust deposition, for the last 17,000 years. c) dust fluxes at the two sites near the Bahamas, OCE205-2 100GGC (26.0612°N, 78.0277°W) and 103GGC (26.0703°N, 78.05617°W), for the last 22,000 years. d) dust flux measured at the VM20 site (5.33°N, 33.03°W) for the last 20,000 years. e) The northern-southern hemisphere temperature difference for the last 17,000 years. The M13[22] uses temperature difference between the extratropics (90 and 30) of two hemispheres and the S12[23] uses full hemisphere averages.

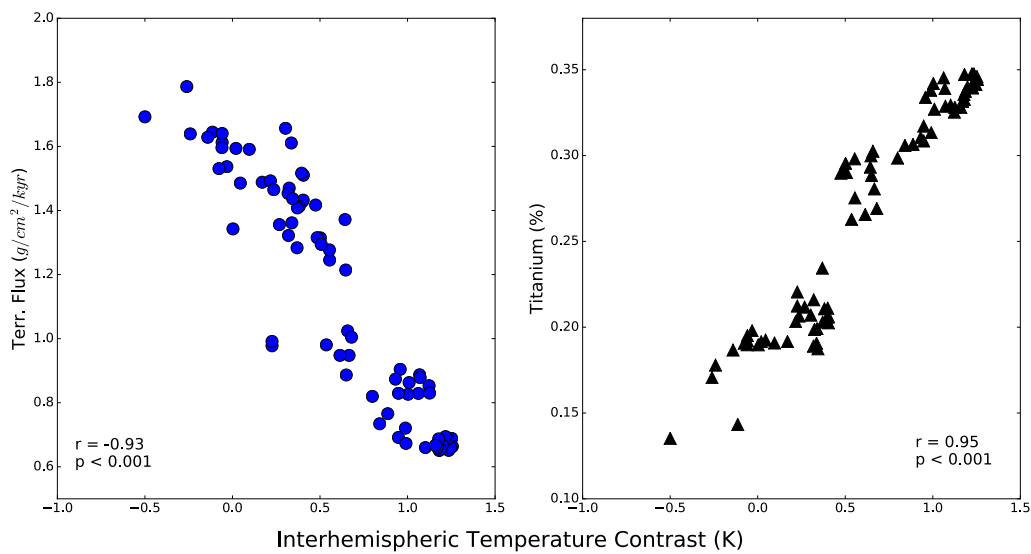

Figure 4: Scatter plot between ICAS proxy and terrigenous flux at ODP658C (left) and the ITCZ position proxy (right) for the last 11,300 years. Both correlations are statistically significant and suggest robust relationships for the last 11,300 years. Each point represents a terrigenous flux measurement and corresponding ICAS and Titanium records averaged over the same period that the dust record covers. Similar evidence is found for the period between 6,000BP and 17,000BP (SOM) using data from Ref[23] (see Fig. S2).

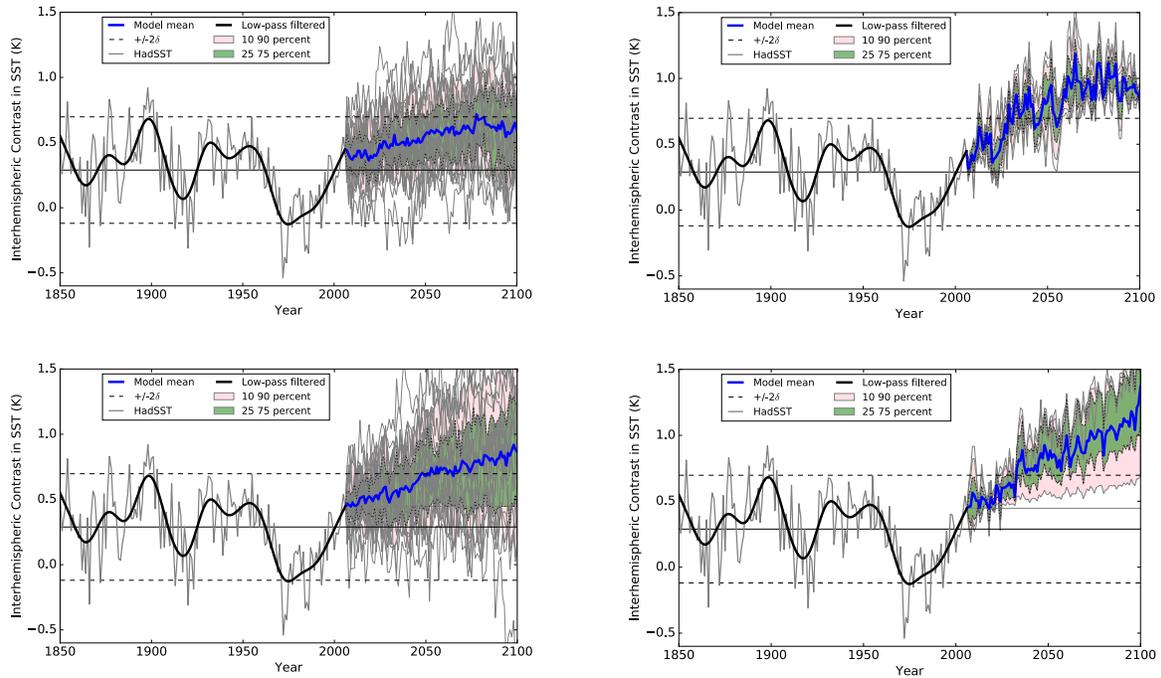

Figure 5: The HadISST based ICAS splined with model simulated ICAS. The horizontal solid lines mark the mean ICAS during the last 167 years. The dashed lines are two standard deviations from the mean. Multi-model means for RCP4.5 (upper panels) and RCP8.5 (lower panels) are plotted using the solid blue lines. All (left) and only "good" (right) models are included (see Text and related reference for details). The criteria for "good" models are given in Data and Method. The green shading marks the 25% to 75% range of the distribution among models and the pink the 10% to 90% range. Individual model results are plotted using solid grey lines.

**Data and Method**

**Paleo-records matching**

The paleo temperature record, Cariaco Basin Ti concentration, and ODP658C terrigenous flux records all have different resolutions. To perform linear regressions, a pair of variables are matched in time: The higher resolution record is reduced to resolution of the lower resolution record by averaging data points in the higher resolution record whose time stamps are closest to a lower resolution data point.

**Reanalysis data**

The ERA-interim reanalysis surface wind data between 1979-2015 are used to calculate regression map of the ICAS against surface wind speed.

**Precipitation data**

The Global Precipitation Climatology Project[35] (GPCP) v2.3 data are used to calculate regression of the ICAS against precipitation.

**CMIP5 data**

Table 1 lists the models used in RCP4.5 and RCP8.5 scenarios. For every model, we follow the same procedure to calculate the ICAS based on model SST. The model data start at 2006 and each model ICAS is adjusted so that its 2006 value is equal to the 2006 value based on observations. Multi-model mean of future ICAS is simply the mean of all adjusted participating models' simulated future ICAS.

Table 1: Models used in this study for RCP4.5 and RCP8.5 scenarios.

| Model Name | RCP4.5 | RCP8.5 |
|---|---|---|
| ACCESS1-0 | ✓ | ✓ |
| ACCESS1-3 | ✓ | ✓ |
| CanESM2 | ✓ | |
| CCSM4 | ✓ | |
| CMCC-CM | ✓ | ✓ |
| CMCC-CMS | ✓ | ✓ |
| CNRM-CM5 | ✓ | ✓ |
| FGOALS-s2 | | ✓ |
| GFDL-CM3 | ✓ | ✓ |
| GFDL-ESM2G | ✓ | ✓ |
| GFDL-ESM2M | ✓ | |
| GISS-E2-R | ✓ | ✓ |
| HadGEM2-AO | ✓ | ✓ |
| HadGEM2-ES | ✓ | ✓ |

| Model name | | |
|---|---|---|
| INMCM4 | ✓ | ✓ |
| MIROC5 | ✓ | ✓ |
| MIROC-ESM-CHEM | ✓ | ✓ |
| MIROC-ESM | ✓ | ✓ |
| MPI-ESM-LR | ✓ | ✓ |
| MPI-ESM-MR | ✓ | ✓ |
| MRI-CGCM3 | ✓ | ✓ |
| MRI-ESM1 | | ✓ |

Table 2 lists the models that are included in the 'poor' and 'good' model categories based on criteria used in[32]. Briefly, good models are those that simulate large AMO amplitude and have good correlation between AMO and Sahel precipitation in the model fields. Results for the poor models can be found in Fig. S3.

Table 2: Models used in the poor and good categories.

| Model name | Good | Poor |
|---|---|---|
| MIROC-ESM-CHEM | ✓ | |
| HadGEM2-ES | ✓ | |
| MIROC5 | ✓ | |
| CNRM-CM5 | | ✓ |
| MPI-ESM-LR | | ✓ |
| CanESM2 | | ✓ |

**Proxy for ITCZ Position**
Titanium concentration data from the anoxic Cariaco Basin (~10N, 65W) have been shown to correlate well with the Atlantic ITCZ latitudinal position and thus provide a good proxy for the ITCZ position[24]. Using Titanium concentration as an ITCZ position proxy also agrees with ITCZ proxy data from other locations around the world and has a physical interpretation[36]: The site sits on the northern edge of the annual ITCZ position and when the ITCZ moves northward, upstream river flow increases as a result of increased winter time precipitation, which brings more titanium and iron to the basin, thus creating a positive correlation between ITCZ position and the Titanium concentration. At the same time, trade wind speed slows down over the ocean and decreases the upwelling, which diminishes the biogenic material input to the basin. All of these connections manifest in the Cariaco Basin record.

**Proxy for ICAS**
At time scales of hundreds of years or longer, we assume that temperature difference between northern hemisphere mean and southern hemisphere mean can be used to represent ICAS since it is found that ICAS drives most of the global interhemispheric temperature contrast[22,23]. The main data source used here comes from the study by Marcott et al.[22], referred to as M13. Northern hemisphere average for the extratropics (30N~90N) and its counterpart in the southern hemisphere are used. This record spans the last 11,300 years. Most of this temperature data comes from maritime records and the extratropical Atlantic signal dominates with minimum contribution from the tropics. Data from the study (Ref[23]), referred to as S12, are also used that covers between 6000BP and 17,000 BP. They are temperature differences between the northern hemisphere mean and the southern hemisphere mean. Because of the spatial coverage difference, 1K is added to the S12 data so that two data records are in excellent agreement during the overlapping period.

**Dust observations and proxies and SPI**

At a Barbados island, surface dust mass concentration has been measured since 1965 by the research group led by Dr. Prospero[6]. The station is located at 13º10' N 59º 30' W about 5000km from the west African coast.

The Cape Verde record is based on reconstruction of $He^4$ measured from a coral at Cape Verde and comparison of this measurement with AVHRR and MODIS satellite data (Evan et al., 2010). It gives a record of dust concentration proxy at 14º55' N 23º 31' W, which is extrapolated to cover the tropical North Atlantic based on satellite data. It runs between 1955 and 2008.

The ERA-I and CIRES-20CR data are from Evan et al., 2016. It is based on projecting ERA-I and CIRES-20CR surface wind speed data onto the second mode of the empirical orthogonal function analysis of the ERA-I surface wind speed over the Northern Africa that covers the African dust source areas.

Paleo dust proxy records for the Bahamas sites and the tropical Atlantic site are from Williams et al.[37]. Dust records from the ODP658C site is from McGee et al. (2013)[38].

The Sahel precipitation index is from the Joint Institute for the Study of the Atmosphere and Ocean at the University of Washington (http://research.jisao.washington.edu/data_sets/sahel/). Its digital object identifier is doi:10.6069/H5MW2F2Q. It is based on station data only over a region that is picked out by a rotated principal component analysis of the African station precipitation data.

**Acknowledgement** We are grateful to the data providers including J. Prospero, A. Evan, D. McGee, R. Williams, G. Haug, J. Shakun, and S. Marcott for sharing their datasets. This effort of T. Y. was supported by the National Aeronautics and Space Administration (NASA) Modeling Analysis and Prediction program (NNX13AM19G).

**Author Contributions:** T.Y. conceived the idea and carried out the bulk of analysis. H.Y. analyzed the Barbados surface data. T. Y. wrote the manuscript with major inputs from H.Y., M.C., and L.A.R. Correspondence and requests for materials should be made to T. Y.